\documentstyle[prd,aps]{revtex}

\newcommand{\be}{\begin{equation}}
\newcommand{\ee}{\end{equation}}
\def\bq{\begin{eqnarray}}
\def\eq{\end{eqnarray}}
\def\th{\theta}

\begin{document}

\title{A curious electrovac spacetime as $G = 0$ limit of the NUT space}
\author{Naresh Dadhich \footnote{e-mail: nkd@iucaa.ernet.in}}
\address{Inter-University Centre for Astronomy and Astrophysics,\\
Post Bag 4, Ganeshkhind, Pune-411 007, INDIA.}
\author{L K Patel}
\address{Department of Mathematics, Gujarat University, Ahmedabad
 380 009, INDIA.}

\maketitle

\begin{abstract}

We take the $G = 0$ limit of the NUT space which yields a non flat space and  
show that source of its curvature is electromagnetic field generated by the 
NUT parameter defining the NUT symmetry. This is a very curious electrovac 
NUT space. Further it is also possible to superpose on it a global monopole. 

\end{abstract}

\vspace{1cm}

It is quite well-known that the NUT solution describes gravomagnetic effects 
of gravity and it could in fact be interpreted as describing  gravitational 
field of a gravomagnetic 
charge [1-5]. The solution has certain unfamiliar features such as it is 
unlike other isolated source solutions is not asymptotically flat. At any 
rate it is an interesting exact solution of the Einstein vacuum equation. 
Its elctrovac generalization is also known [6] which describes a NUT particle 
with electric charge. 

For asymptotically flat spacetimes, the limit of vanishing gravitational 
constant, $G = 0$ yields flat space\footnote{A very simple and novel electromagnetic field has 
been derived as $G = 0$ limit of the charged rotating Kerr-Newman metric [8].}. It should be interesting to find 
what does this limit give for asymptotically non flat case of the NUT 
solution? This is what we wish to explore in this note. It would turn out that 
source of curvature of the $G = 0$ limit of the NUT space is interestingly 
the Maxwell field generated by the NUT parameter. The NUT parameter occurs in 
the metric in two roles, one in 
gravitational potential and the other in defining the symmetry of the space. 
In the $G = 0$ limit, the former role washes away while the latter still 
survives and it is this that produces the electromagnetic field which is 
responsible for the curvature. It is remarkable that it is the pure NUT space 
which is endowed with electric charge which produces no gravitational 
potential.

The NUT space is described by the metric,
\be 
ds^2 = A(dt - 2l\cos\th d\varphi)^2 - A^{-1}dr^2 - (r^2+l^2)(d\th^2 + 
\sin^2\th d\varphi^2)
\ee
with
\be
A = 1 + 2\phi, \, ~ \phi = -G\frac{Mr + l^2}{r^2 + l^2}.
\ee 
Here $\phi$ is the gravitational potential of the NUT particle of mass M and 
the NUT charge $l$. 

Now we take the $G = 0$ limit, which is equivalent to gravitational potential 
vanisning, and so the metric reduces to 
\be 
ds^2 = (dt - 2l\cos\th d\varphi)^2 - dr^2 - (r^2+l^2)(d\th^2 + \sin^2\th 
d\varphi^2).
\ee
It could be easily verified that it is not flat and the source of its 
curvature is electromagnetic field given by
\be
{\bf F} = \frac{\sqrt{2}l(r^2 - l^2)}{(r^2 + l^2)^2}{\bf d}r \wedge ({\bf d}t 
- 2l\cos\th {\bf d}\varphi)
\ee
which generates in view of the Einstein vacuum equation $R_{ab} = 
T_{ab}$ the following stresses,
\be
T^0_0 = T^1_1 = -T^2_2 = -T^3_3 = -\frac{2l^2}{(r^2 + l^2)^2}
\ee
where we have set $8\pi G = 1 = c$.

It is pure NUT space which gets endowed with an effective electric 
charge, $\sqrt{2}l$ which produces a radial electric field with effective 
squared area radius, $r^2 + l^2$. The above energy distribution obviously 
satisfies the positive energy conditions. It is a very curious and amazing 
spacetime which is free of the Newtonian gravitational potential. Radial 
motion is indistinguishable from that of flat space and else all the 
geodesics lie on spatial cones [5] with the semi-angle determined by $l$ and 
angular momentum of the test particle. 

The electromagentic field vanishes at $r = l$ and asymptotically it falls off 
as $1/r^2$. It attains the maximum value $1/{4\sqrt{2}l}$ at $r^2 = 3l^2$. In 
the Kerr - Newman case, it vanishes at $r = a\cos\th$ (which would always lie 
inside the horizon) and maximum occurs at $r^2 = 3a^2\cos^2\th$. The field 
is singular on the ring $r=0, \th = \pi/2$, which is also the curvature 
singularity. In the present case, the spacetime neither admits 
horizon nor does it have singularity of any kind anywhere. It is regular 
everywhere and but it does not go flat asymptotically. It appears as though 
the surface $r = l$ is a perfect conductor and carrying charge $\sqrt{2} l$. 
Asymtotically, it would appear as the usual Coulombic field.
 
Recently, it has been attempted to identify the physical symmetry of the NUT 
space [7] as the gauged Killing symmetry. For the $1 + 3$ decomposition of 
spacetime, the metric would have the block, $h(dt + A_a dx^a)^2$ where $A_a$ 
does behave like a gauge vector potential for gravomagnetic part of the 
field [5]. The gauged Killing symmetry is defined by
\be
\L_\xi g_{ab} = h \psi_{;(a} A_b)
\ee
where
\be
\L_\xi A_a = \psi_{;a}.
\ee
Obviously the NUT solution belongs to this class and it can be verified that 
it admits the gauged Killing symmetry, which is characteristic of the NUT 
space. 

Next, let us write in the metric (1), $A = 1 + 2k$, where $k$ is a constant. 
This would correspond to constant gravitational potential in the NUT space. We 
have seen above that pure NUT space with vanishing gravitational potential 
gets endowed with electromagnetic field. On the other hand, it has been shown 
that constant potential superposes a global monopole field [9-12] on the 
existing space. Inclusion of constant potential would lead to the following 
stresses,
\bq
T^0_0 &=& -\frac{2l^2(1 + 2k)^2}{(r^2 + l^2)^2} + \frac{4k(1 + k)}{r^2 + 
l^2} = T^1_1, \\
T^2_2 &=& T^3_3 = \frac{2l^2(1 + 2k)^2}{(r^2 + l^2)^2}.
\eq
A global monopole is supposed to be produced in phase transition in the 
early universe by instantaneous breaking of global $O(3)$ symmetry into 
$U(1)$. Global monopole on a NUT particle of the parameters $M, l$ has been 
studied [13]. It is modelled as a triplet scalar field, $\psi^a = \eta f(r)
x^a/(r^2 + l^2)^{1/2}, x^ax^a = r^2$ with the usual scalar field Lagrangian 
[13, 9, 11, 12]. The stresses it generates would asymptotically approximate to 
$T^0_0 = T^1_1 = \eta^2/(r^2 + l^2)$ and all others vanishing. Clearly the 
above stresses show that there is superposition of the original 
electromagnetic field and global monopole field with the identification 
$\eta^2 = 4k(1 + k)$.  
   
In the above metric, gravitational source is purely the gauge potential 
$A_a$ and hence it could cause no radial acceleration as indicated by the 
absence of the Newtonian gravitational potential. The parameter $l$ 
essentially defines the gauged Killing symmetry and it is remarkable that 
it also sources the electromagntic field as given above. This brings out the 
intimate relation between the NUT gauged symmetry and electromagnetic field. 
That is electromagnetic field is kind of endowed in the symmetry itself. 
Gravoelectric charge (mass/energy distribution) defines gravitational 
potential and does not directly enter into defining symmetry of space. Here 
it primarily defines the symmetry and in turn it gives rise to electric field 
which is the cause of curvature of the metric. 

Further we have shown that it is also possible to superpsoe global monopole 
field, as is the case for general spherically symmetric spacetime [12], on 
this pure NUT space. It is interesting to note that in the NUT space when we 
make gravitational potential vanish we end up with a spacetime  generated by 
an electromegnetic field, and when potential is constant then a global 
monopole gets superposed on the original electromagnetic source. Of course 
when 
gravitational potential is not tampered with, then it is the vacuum NUT 
solution. However, it is a remarkable and novel way of generating an 
electromagnetic source. The association of NUT gauged symmetry [7] with  
electromagnetic field calls for further probing and it may perhaps reveal 
something important and insightful.

{\bf Acknowledgement:} LKP would like to thank IUCAA for the visit during 
which this work was done. It is a pleasure to thank Kesh Govinder for 
computational help.

\end{document}